\documentclass[a4paper]{jpconf}
\usepackage[utf8]{inputenc}
\usepackage{amsmath}
\usepackage{amsfonts}
\usepackage{amssymb}
\usepackage{graphicx}
\usepackage{placeins}
\usepackage{color}
\usepackage{textgreek}

\begin{document}
	\title{Readout system and testbeam results of the RD50-MPW2 HV-CMOS pixel chip}
	\author{P Sieberer$^1$, T Bergauer$^1$, K Flöckner$^1$, C Irmler$^1$, H Steininger$^1$} 
\address{$^1$ Institute of High Energy Physics (HEPHY), Austrian Academy of Sciences (OEAW), Nikolsdorfer Gasse 18, 1050 Wien, AT}
\ead{patrick.sieberer@oeaw.ac.at}

\begin{abstract}
	The RD50-CMOS group aims to design and study High Voltage CMOS (HV-CMOS) chips for use in a high radiation environment. Currently, measurements are performed on RD50-MPW2 chip, the second prototype developed by this group. 
	The active matrix of the prototype consists of 8x8 pixels with analog front end. Details of the analog front end and simulations have been already published earlier. This contribution focuses on the Caribou based readout system of the active matrix. Each pixel of the active matrix can be readout one after the other. Relevant aspects of hardware, firmware and software are introduced. As a first stage, firmware for a standalone setup is introduced and details on data flow are given. Afterwards, a second stage of the firmware capable of synchronizing with other detectors and accepting triggers is presented, focusing on operation of the chip in combination with a tracking telescope to measure efficiency and residuals. 
\end{abstract}

\FloatBarrier
\section{Introduction}
Depleted monolithic active pixel sensors (DMAPS) are a promising option for future tracking detectors in high energy physics. The availability of high-resistivity Si substrates and High Voltage (HV) capable structures in commercial CMOS processing makes them suitable as detectors in high-luminosity environments. DMAP sensors integrate the sensing diode into the bulk silicon of commercial CMOS chips. It also allows to use very small transistors which results in high granularity pixels with sophisticated in-pixel logic. Thus, amplification, sampling and digitization of signals as well as timing measurement and data reduction can be done on-chip.

The RD50 HV-CMOS group is studying radiation-hard DMAP sensors in the course of the CERN RD50 collaboration.
This includes not only sensor development itself, but also characterization, electronics for data acquisition (DAQ) and readout software. This contribution focuses on the DAQ system and data flow of the RD50-MPW2 chip.

\FloatBarrier
\section{RD50 HV-CMOS Road Map}
The first chip developed by the RD50 HV-CMOS group was manufactured in a multi project wafer (MPW) run and called RD50-MPW1. The chip was fabricated in the LFoundry 150nm process with a wafer thickness of 280\textmu m. 
Measurements have shown, that the analog behavior of RD50-MPW1 was suffering from high leakage current and low breakdown voltage~\cite{mpw1mpw2}. Moreover the chip showed some shortcomings in the digital readout. 
In order to tackle one issue after the other, the group decided to focus on improvements of the analog behavior of the pixel cell and designed a second iteration of the chip, the RD50-MPW2. The chip only has an analog front end in the pixels and no digital readout. As a result of these improvements, the pixel in the active matrix have been enlarged from 50\textmu m x 50\textmu m in RD50-MPW1 to 60\textmu m x 60\textmu m in RD50-MPW2, still focusing on the minimal possible size to achieve a highly granular matrix. 
An enlarged distance between the collection electrode and the biasing implant lead to a decrease of leakage current by four orders of magnitude. This massive improvement is suitable to be used in future tracking devices, where power dissipation is one of the big challenges. The analog characteristics and performance have been presented for example in~\cite{vertex2020}.

In the course of characterizing and testing of RD50-MPW1 and later the RD50-MPW2 chips, a readout chain to be used in laboratory measurements as well as for beam tests has been implemented. In addition to the standalone data acquisition, this readout system provides techniques to synchronize with other detectors by an AIDA trigger logic unit (TLU)~\cite{tlu}.


\FloatBarrier
\section{Overview RD50-MPW2} 
\begin{figure}[!ht]
\begin{minipage}[c][][b]{0.48\textwidth}
	\begin{center}
		\includegraphics [width=0.8\textwidth]{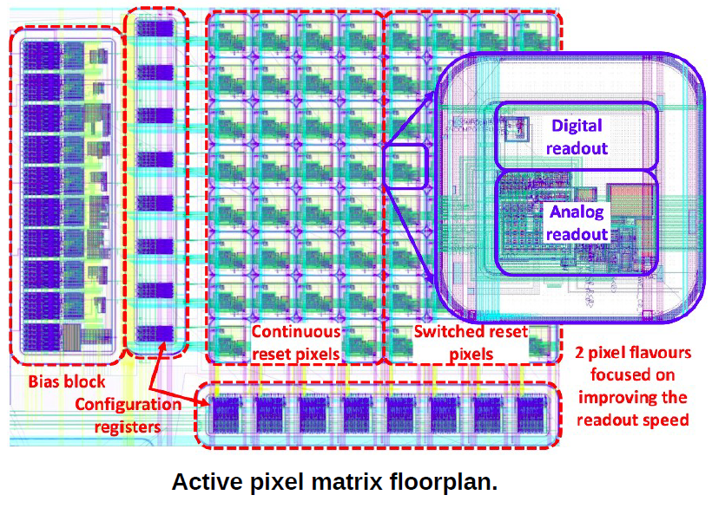}
		\caption{\label{fig:matrix} The analog pixel matrix of RD50-MPW2~\cite{mpw2doc}.}
	\end{center}
\end{minipage}
\hfill
\begin{minipage}[c][][b]{0.48\textwidth}
	\begin{center}
		\includegraphics [width=\textwidth]{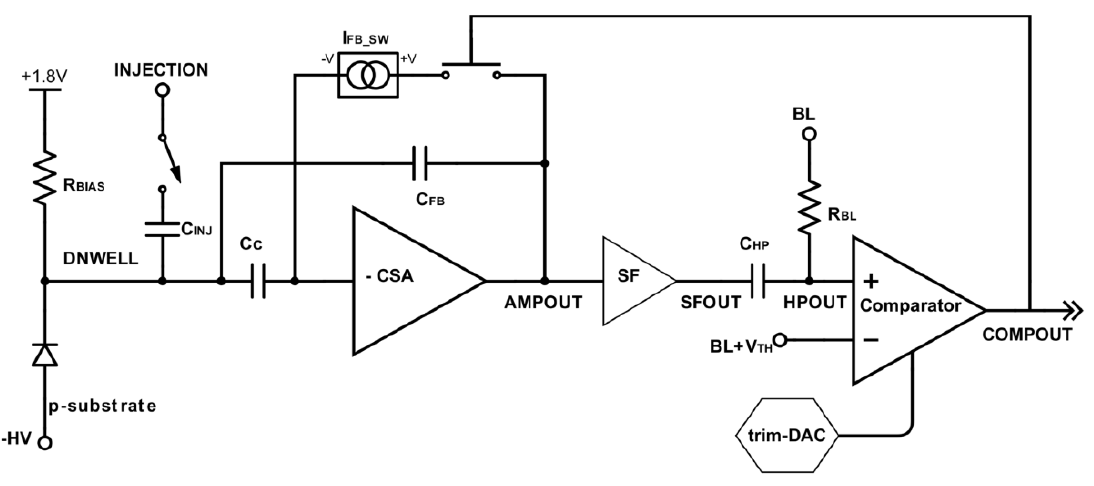}
		\caption{\label{fig:switchedPixelSchematics} Schematic of a switched reset pixel~\cite{mpw2doc}.}
	\end{center}
\end{minipage}
\end{figure}

RD50-MPW2 consist of different test structures for studying the analog behavior of the chip~\cite{mpw2doc}. One main part is the active pixel matrix, which is depicted in figure \ref{fig:matrix}. It includes a bias block, configuration registers (implemented as shift register) and the pixel matrix itself with 8 by 8 pixels and a size of 60\textmu m x 60\textmu m each. Each pixel has a charge sensitive amplifier (CSA), an analog output after a source follower, called SFOUT, and a digital output after a comparator, called COMPOUT, as shown in Figure~\ref{fig:switchedPixelSchematics}.
The matrix provides pixels of two different flavors of readout: columns 0-3 are continuous reset pixels, while columns 4-7 are switched reset pixels. 
They differ in their feedback loop of the CSA. The feedback capacitor of the continuous reset pixel is continuously discharged by a fixed current $\rm I_{FB\_CONT}$ source, leading to an amplitude dependent output signal length. For the switched reset pixel, the feedback capacitor is discharged by a comparably high current $\rm I_{FB\_SW}$ source, which is switched on as soon as the signal is above a threshold and COMPOUT is high, resulting in a charge independent pulse length.
A pixel from the latter type is chosen for all measurements in this contribution.

It is important to mention that the matrix is analog only and moreover only one pixel at a time can be read out. The output of each pixel is tri-stated and ORed together for each column and all columns are again ORed together and routed to a bonding pad. This is done for both SFOUT and COMPOUT, respectively.

\FloatBarrier
\section{Standalone DAQ for laboratory tests}\label{sec:standalone}
The DAQ system is based on the Caribou framework \cite{caribou}, which includes hardware, firmware and software packages for configuring, powering and reading out pixels sensors. A detailed description of the setup used for RD50-MPW2 can be found in \cite{chrisuCaribou}. This contribution focuses on firmware developments.

\begin{figure}[!ht]
\begin{center}
	\includegraphics[width=0.8\textwidth]{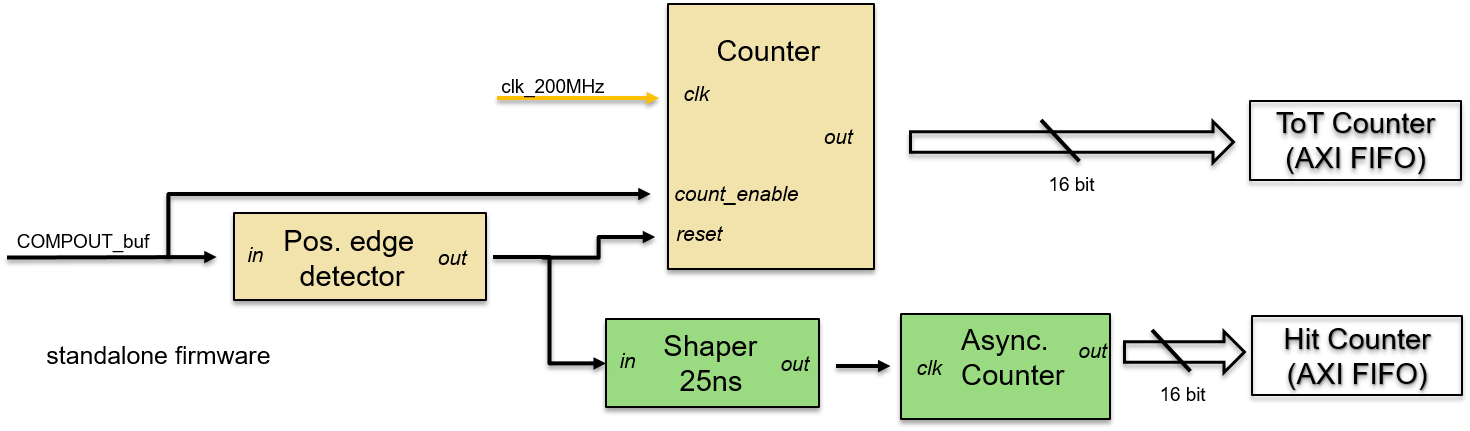}
\end{center}
\caption{\label{fig:standaloneFW} Standalone firmware of RD50-MPW2.}
\end{figure}

Figure \ref{fig:standaloneFW} shows a schematic of the standalone firmware, which can be used for laboratory measurements such as charge injection by the integrated injection circuit and charge generation by a radioactive source. The COMPOUT is buffered and connected to the FPGA (located at the Xilinx SoC ZC706 Evaluation Board, which is part of Caribou). This simple implementation includes a counter with a 200MHz clock input which is set and reset at the two edges of the pulse. It provides the 16-bit time over threshold (ToT) value at the output. Furthermore, one edge is shaped and counted in an asynchronous counter, which provides another 16-bit as output for counting particles. Those two times 16 bits are feed into the AXI FIFO, which is part of the standard AXI-bus from Xilinx \cite{axi}.

The functions to access these FIFOs with the CPU are already implemented in the Caribou framework.
The readout software just polls the AXI FIFOs every 100\textmu s and writes the values into an ASCII file. 
The integrated peary command-line interface (peary\_cli) is used for this purpose.

\FloatBarrier
\section{Testbeam setup and DAQ}
\subsection{Synchronizing with other detectors}\label{sec:testbeam}
\begin{figure}[!ht]
\begin{center}
	\includegraphics[width=0.8\textwidth]{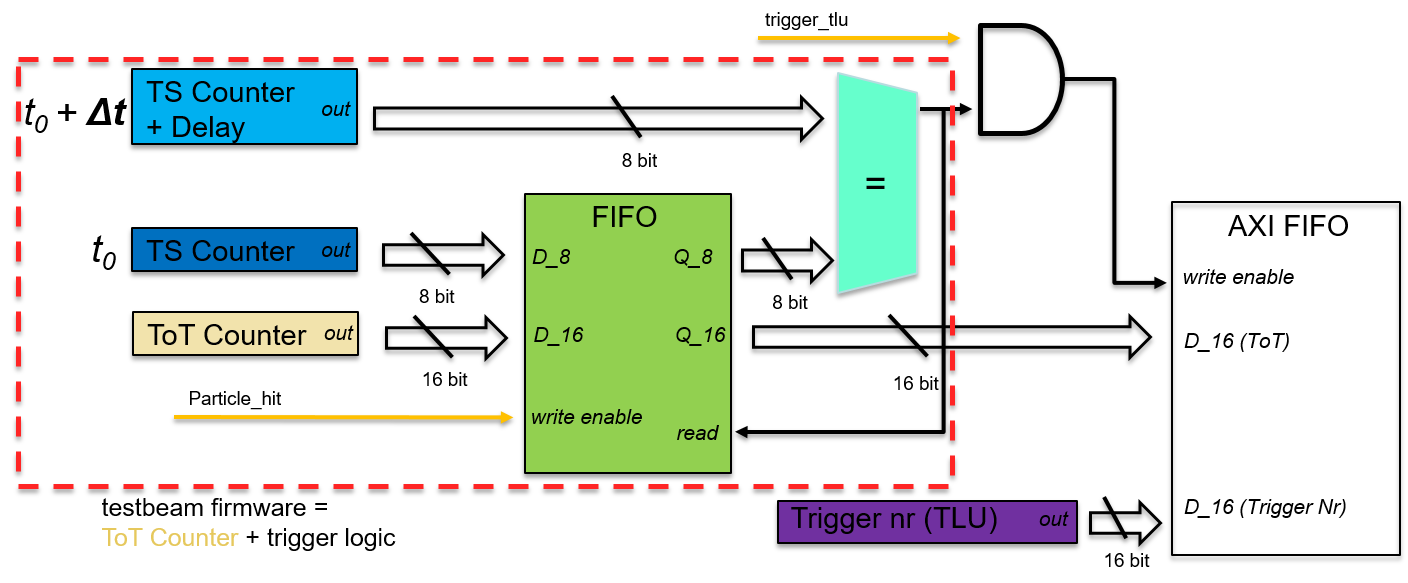}
\end{center}
\caption{\label{fig:testbeamFW} Testbeam firmware of RD50-MPW2.}
\end{figure}

Synchronization with other detectors and the capability of accepting trigger signals is a crucial need of a detector in order to work in a tracking setup during testbeams. Thus, the standalone firmware was modified to implement this functionality. The extended firmware is called testbeam-firmware and a schematic is depicted in figure \ref{fig:testbeamFW}. The main component is another FIFO which has a delayed input.

The FIFO captures the ToT value, calculated in the same way as in the standalone-firmware, whenever a particle hits the collection electrode. The second input of the FIFO is an internal counter. The FIFO is read out whenever the value of the counter matches with another counter working at the same speed, but delayed by a certain amount of time \textDelta t. The blocks implementing this behavior are enclosed by the red-dashed rectangle in figure \ref{fig:testbeamFW}.

The whole circuit has another input, the trigger signal, which is generated by the TLU. It is issued by the coincidence of two scintillators placed behind the detector and only forwarded to all sub-detectors if none of them is vetoing triggers. This is needed, in order to record data only when every detector is ready. Other data is thrown away, which is done in RD50-MPW2 with a simple AND gate which enables the writing of the AXI FIFO. If a trigger is issued, the (delayed) value of ToT is written into the AXI FIFO along with a trigger number from the TLU in order to synchronize events with other detectors. The important parameter of this setup is the time of the delay \textDelta t, which needs to match exactly with the trigger delay. This was determined before the testbeam by directly measuring \textDelta t with an oscilloscope.

\FloatBarrier
\subsection{Testbeam setup and beam parameters}
Both setups where tested with a proton beam at MedAustron~\cite{Benedikt:67546}. Since this is a medical facility for cancer treatment with a high nominal particle rate in the GHz region, so-called low-flux parameters have been determined in order to provide a more suitable beam for tracking devices~\cite{lowflux}. For testing RD50-MPW2, setting "high (MED)" at an energy of 252.7\;MeV and a proton rate of 3.73 - 5.21MHz was used. Due to restrictions by the telescope, the overall event rate was at around 30\;kHz.

\FloatBarrier
\section{Testbeam results}
\begin{figure}[!ht]
\begin{minipage}[c]{0.48\textwidth}
	\begin{center}
		\includegraphics [width=\textwidth]{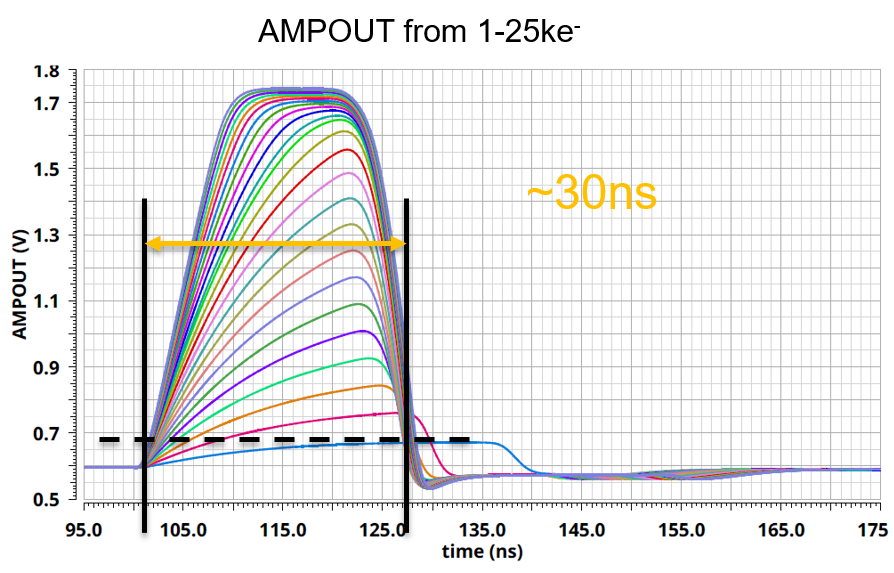}
		\caption{\label{fig:tot_simu} Simulation of the analog output gives a ToT of about 30ns \cite{mpw2doc}.}
	\end{center}
\end{minipage}
\hfill
\begin{minipage}[c]{0.48\textwidth}
	\begin{center}
		\includegraphics [width=\textwidth]{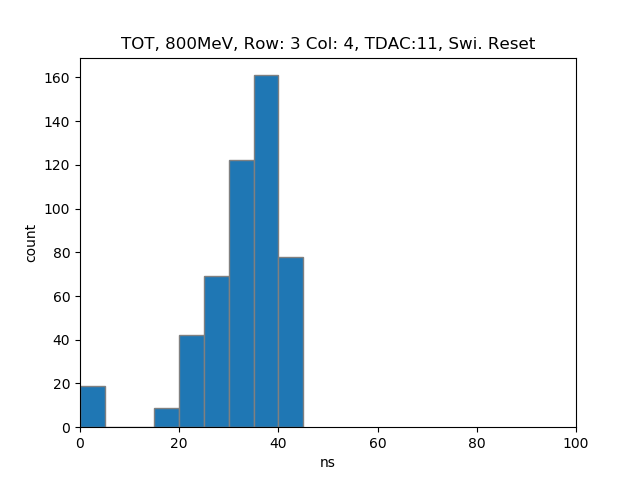}
		\caption{\label{fig:tot} ToT measurements using the testbeam-firmware show a distribution peaking at 35ns.}
	\end{center}
\end{minipage}
\end{figure}

In order to test the standalone firmware RD50-MPW2 serves as device under test (DUT) and is put directly into the beam. As one can see in figure~\ref{fig:tot_simu}, the ToT value is independent from deposited energy for switched reset pixels, since the CSA is resetting nearly instantaneous after the switch is closed.
The standalone firmware of RD50-MPW2 roughly measures this value as well, but with a bin size of 5\;ns, which is needed for measuring longer pulses from the other pixel flavour. Calibration to get charge in units of electrons has not been performed yet.

In order to test the testbeam-firmware, RD50-MPW2 is put in the middle of a beam telescope, which uses two double sided silicon strip detectors (DSSDs)  upstream and two DSSDs downstream of the DUT. The telescope is triggered by the coincidence of two scintillators, which are located at the very back of the setup. Analysis of the tracking capability of RD50-MPW2 is ongoing. Due to small geometric acceptance (just one single pixel is recorded at the same time) very high statistics is needed to get reliable results for tracking studies. This is in contradiction to the maximum rate of the telescope of 30\;kHz.
Thus the goal of this setup is to measure pixel efficiency, which can be done with a single pixel. 

\begin{figure}[!ht]
\begin{minipage}[c]{0.5\textwidth}
	\begin{center}
		\includegraphics[width=\textwidth]{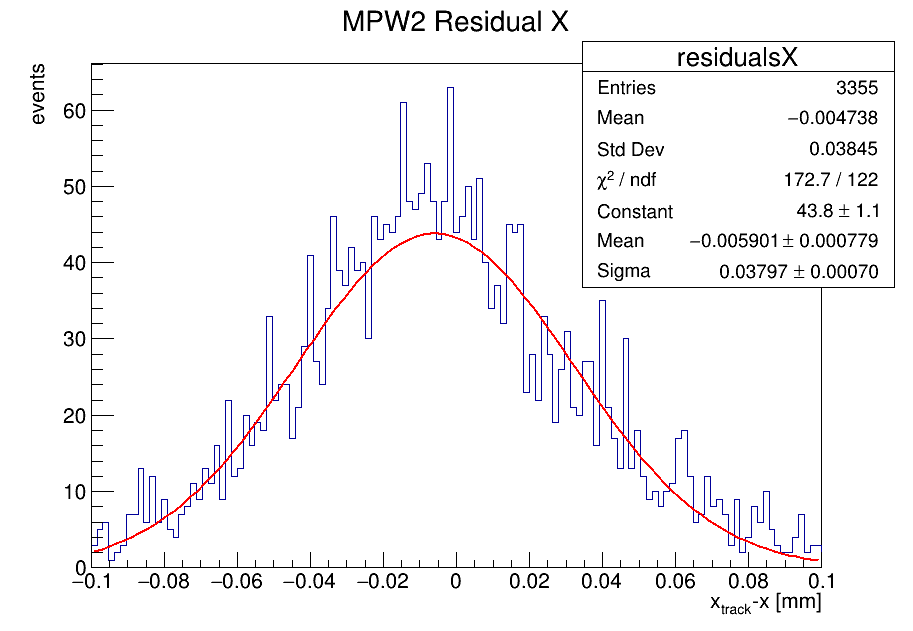}
	\end{center}
\end{minipage}
\hfill
\begin{minipage}[c]{0.48\textwidth}
	\caption{\label{fig:resDUT} \footnotesize The depicted distribution of residuals for the DUT are centered around 0 and have a distribution very similar to the planes of the telescope. This is expected, since only a constant value, the position of the pixel, is subtracted from the uncertainty of the beam position, which has no influence on the distribution of residuals. The rather low beam energy causes multiple scattering of beam particles in air, which widens the distribution of residuals for all detectors. It is only in the order of half of a pitch for the telescope planes, instead of the theoretically expected value ${\it pitch}/\sqrt{12}$.}
\end{minipage}
\end{figure}


\FloatBarrier
\section{Conclusion and outlook}
In this paper we report the first data taking with the RD50-MPW2 DMAPS chip in a testbeam environment.
It has been shown that both readout firmware versions work very well and demonstrate a good agreement of the analog performance of the chip with simulation results. Also the synchronization of the chip with other detectors is working and analysis of tracking data shows consistent results.
The main goal of this measurement was to familiarize with the special high-rate, but low-energy environment at MedAustron and prepare the DAQ system and analysis software for the successor chip of RD50-MPW2, which has been achieved.

Currently, RD50-MPW3 is being developed which will have a digital front end and the capability of reading out the full matrix. Knowledge gained for RD50-MPW2 in DAQ and analysis will be a valuable input for the study of this successor sensor.

\ack
This work was performed in the framework of the CERN RD50 collaboration. This project has received funding from Austrian Research Promotion Agency (FFG), grant number 878691.

\section*{References}		
\bibliography{tipp_bib}


\end{document}